\documentclass[useAMS,usenatbib]{mn2e}

\newcommand{\age}{t_{\rm age}}
\usepackage{amsmath, amssymb}
\usepackage[dvips]{graphicx}

\usepackage{color}
\usepackage{ulem}

\title[Properties of Young PWNe]{Properties of Young Pulsar Wind Nebulae: TeV Detectability and Their Pulsar Properties}

\author[S. J. Tanaka and F. Takahara]{Shuta J. Tanaka$^{1}$\thanks{E-mail: tanaka@phys.aoyama.ac.jp} and Fumio Takahara$^{2}$ \\
$^{1}$Department of Physics and Mathematics, Aoyama Gakuin University, 5-10-1 Fuchinobe, Sagamihara, Kanagawa, 252-5258, Japan \\
$^{2}$Department of Earth and Space Science, Graduate School of Science, Osaka University, 1-1 Machikaneyama-cho, Toyonaka, Osaka 560-0043, Japan
}

\begin{document}

\date{Accepted 1988 December 15. Received 1988 December 14; in original form 1988 October 11}

\pagerange{\pageref{firstpage}--\pageref{lastpage}} \pubyear{2012} 
\maketitle

\label{firstpage}

\begin{abstract}
Among dozens young pulsar wind nebulae, some have been detected in TeV $\gamma$-rays (TeV PWNe), while others have not (non-TeV PWNe).
The TeV emission detectability is not correlated either with the spin-down power or with the characteristic age of their central pulsars, and it is an open problem what determines the detectability.
To study this problem, we investigate spectral evolution of five young non-TeV PWNe, 3C58, G310.6-1.6, G292.0+1.8, G11.2-0.3 and SNR B0540-69.3.
We use a spectral evolution model which has been developed to be applied to young TeV PWNe in our previous works.
TeV $\gamma$-ray flux upper limits of non-TeV PWNe give upper or lower limits on parameters, such as the age of the PWN and the fraction of the spin-down power going to the magnetic energy injection (the fraction parameter).
Combined with other independent observational and theoretical studies, we can guess a plausible value of the parameters for each object. 
For 3C58, we prefer the parameters with an age of 2.5 kyr old and the fraction parameter of $3.0 \times 10^{-3}$, although the spectral modeling alone does not rule out a shorter age and a higher fraction parameter.
The fraction parameter of $3.0 \times 10^{-3}$ is also consistent for other non-TeV PWNe and then the value is regarded as common to young PWNe including TeV PWNe.
Moreover, we find that the intrinsic properties of the central pulsars are similar, $10^{48 - 50} \rm erg$ for the initial rotational energy and $10^{42 - 44} \rm erg$ for the magnetic energy ($2 \times~10^{12} - 3 \times~10^{13} \rm G$ for the dipole magnetic field strength at their surfaces).
The TeV detectability is correlated with the total injected energy and the energy density of the interstellar radiation field around PWNe.
Except for G292.0+1.8, a broken power-law injection of the particles well reproduces the broadband emission from non-TeV PWNe.
\end{abstract}


\section{INTRODUCTION}
Rotation-powered pulsars release most of their rotational energy as a relativistic outflow of a magnetized electron-positron plasma called pulsar winds.
The pulsar wind collides with a surrounding supernova (SN) ejecta and creates a pulsar wind nebula (PWN) \citep[][]{rg74, kc84a}.
The resultant PWN is a cloud of the relativistic magnetized electron-positron plasma dissipated at a termination shock of the pulsar wind and shines from radio through TeV $\gamma$-rays via synchrotron radiation and inverse Compton scattering \citep[][]{kc84b}.

More than fifty PWNe have been identified till now.
Although they share some common observational characteristics, they also show individualities, such as an observed size and a TeV $\gamma$-ray to X-ray flux ratio.
The differences may be due to the properties of the central pulsar (e.g., the spin-down power and the surface magnetic field), of the environment (e.g., SN ejecta, the interstellar medium and the interstellar radiation field (ISRF)) and of the age of each object \citep[e.g.,][]{tt10, tt11, bet11, get09}.
In other words, we can infer the properties of the central pulsars, the environment of the objects and the age of the objects from observed characteristics.

The goal of this paper is to understand which properties of PWNe affect TeV $\gamma$-ray properties.
The ground-based $\gamma$-ray telescopes (e.g., {\it HESS, MAGIC, VERITAS} etc.) have detected TeV $\gamma$-rays from many PWNe (TeV PWNe), but some have not been detected (non-TeV PWNe).
\citet{met09} studied correlations between the TeV $\gamma$-ray luminosity from PWNe and their pulsar properties (the spin-down power and the characteristic age).
However, they found no correlation (c.f., see their Figure 1).
Therefore, we expect that other properties characterise the detectability of TeV $\gamma$-rays from young PWNe.
In this paper, we study this problem from the broadband spectrum of PWNe.

We built a spectral evolution model of young TeV PWNe in our previous works and found that they share common characteristics, such as a magnetic energy fraction of the total energy injected from the central pulsar (TT11).
Here, we apply the same spectral evolution model to young non-TeV PWNe and study their differences and similarities.
After the applications of our model to non-TeV PWNe, we discuss the properties of the whole class of young PWNe.
Note that our spectral evolution model is applicable only to young PWNe which show no signature of an interaction with a reverse shock of a supernova remnant (SNR).
This is because our model does not consider complex dynamical behaviors of PWNe in the interaction phase \citep[e.g.,][]{bet01b, get09}.

We apply the model to five young non-TeV PWNe 3C58, G310.6-1.6, G292.0+1.8, G11.2-0.3 and SNR B0540-69.3 (hereafter B0540-69.3).
We pick up these five PWNe according to four criteria.
(1) They reveal non-thermal spectra at least in radio and X-rays.
(2) They have a flux upper limit in TeV $\gamma$-rays.
(3) They have a central pulsar with a known period and its derivative.
(4) They have an almost spherical shape with a known angular extent, i.e., they are young enough not to reveal any signatures of an interaction with the SNR reverse shock.

In Section \ref{sec:procedure}, we describe the fitting procedure which is different from that for TeV PWNe.
The spectral evolution model is the same as our previous works and is summarized in Appendix \ref{app:model}.
In Sections \ref{sec:3c58} and \ref{sec:other}, we apply the model to the five non-TeV PWNe and present results.
Comparative discussions about young non-TeV PWNe and about young TeV PWNe are made in Section \ref{sec:discussion}.
Conclusions are made in Section \ref{sec:conclusion}.

\section{Fitting Procedure}\label{sec:procedure}

%
\begin{table*}
\begin{minipage}{126mm}
\caption{
Adopted, Fitted, and Derived Parameters of Young PWNe
}
\label{tbl:parameter}
\begin{tabular}{cccccccc}
\hline
Symbol     & TeV PWNe$^{a}$  &
3C58       & 3C58$^{b}$       &
G310.6-1.6 & G292.0+1.8 & G11.2-0.3 &
B0540-69.3 \\
Model & & 1 & 2 & & & & \\
\hline
\multicolumn{8}{c}{Adopted Parameters} \\
\hline

$d$(kpc) &
	2.0  -- 13   & 2.0  & 2.0  & 7.0   & 6.0  & 5.0  & 50    \\
$R_{\rm PWN, now}$(pc) &
	1.0 -- 3.8   & 2.0  & 2.0  & 1.3   & 3.5  & 0.97 & 1.2   \\
$\textit{P}$(msec)  &
	33.1 -- 136  & 65.7 & 65.7 & 31.2  & 135  & 65.0 & 50.7  \\
$\dot{\textit{P}} (10^{-13} \rm sec \cdot sec^{-1})$ &
	1.56 -- 7.51 & 1.94 & 1.94 & 0.389 & 7.47 & 0.440& 4.79  \\
$\textit{n}$ &
	2.51 -- 3.0  & 3.0  & 3.0  & 3.0   & 3.0  & 3.0  & 2.1   \\
$U_{\rm IR}$($\rm eV cm^{-3}$) &
	1.0  -- 2.0  & 0.3  & 0.3  & 0.3   & 0.3  & 1.2  & 0.5   \\
$U_{\rm OPT}$($\rm eV cm^{-3}$) &
	0.5  -- 15   & 0.3  & 0.3  & 0.3   & 0.3  & 2.0  & 0.5   \\

\hline
\multicolumn{8}{c}{Fitted Parameters} \\
\hline

$\eta^{c}$($10^{-3}$) &
1.0 -- 8.0   & 3.0& 60  & 3.0$^{d}$ & 3.0$^{d}$  & 3.0$^{d}$ & 3.0$^{d}$ \\
$\textit{t}_{\rm age}$(kyr) &
	0.95 -- 4.5  & 2.5& 1.0 & 0.6 & 2.7  & 2.0 & 0.7 \\
$\gamma_{\rm max}^{c}$($10^9$) &
	0.8 -- 7.0   & 1.0& 0.5 & 4.0 & 1.5  & 1.0 & 0.8 \\
$\gamma_{\rm b}$($10^5$) &
	0.4 -- 6.0   & 0.9& 0.5 & 30  & 0.3  & 1.0 & 100 \\
$\gamma_{\rm min}^{c}$($10^3$) &
	0.1 -- 20    & 0.4& 0.1 & 5.0 & --   & 2.5 & 4.5 \\
$\textit{p}_1$ &
	1.0 -- 1.5   & 1.0& 1.0 & 1.4 & --   & 1.5 & 1.7 \\
$\textit{p}_2$ &
	2.5 -- 2.6   & 3.0& 3.0 & 3.0 & 2.5  & 2.6 & 2.5 \\

\hline
\multicolumn{8}{c}{Derived Parameters} \\
\hline

$\textit{v}^{}_{\rm PWN}$(km~sec$^{-1}$) &
	770 -- 1800  & 780 &2000 & 2120 & 1270 & 470  & 1680\\
$B_{\rm now}$($\mu \rm G$) &
	6.7 -- 85    & 17  &40   & 17   & 16   & 17   & 40  \\
$\tau_0$(kyr) &
	0.7 -- 3.9   & 2.9 &4.4  & 11.5 & 0.16 & 21   & 2.4 \\
$L_0 \tau_0$($10^{48} \rm erg$) &
	2.6 -- 74    & 8.6 &5.6  & 21   & 19   & 5.2  & 22  \\
$\kappa^{c}$($10^4$) &
	3.4 --  420  & 23  &45   & 3.8  & 5.3  & 6.3  & 9.5 \\
$\Gamma_{\rm w}^{c}$($10^5$) &
	0.07 -- 2.7  & 0.31&1.6  & 1.8  & 0.90 & 4.0  & 1.8 \\
\hline
\end{tabular}

\footnotetext[1]{
Taken from Table 1 of TT11 excluding the models whose $\eta$ are significantly different from a few $\times~10^{-3}$ (model 1 of G21.5-0.9 and G54.1+0.3 and both models of Kes 75).
}
\footnotetext[2]{
The case considering SN 1181 ($\age \sim 1 \rm kyr$).
}
\footnotetext[3]{
The values are upper limits for $\gamma_{\rm min}$ and $\Gamma_{\rm w}$ and lower limits for $\gamma_{\rm max}$ and $\kappa$.
}
\footnotetext[4]{
We fit the broadband spectrum assuming $\eta = 3.0 \times~10^{-3}$ for these objects.
}
\end{minipage}

\end{table*}

For each object, we adopt the distance to the object $d$, the angular extent of the object, and the central pulsar parameters, the period $P$, its derivative $\dot{P}$ and the braking index $n$ from observations.
Then, we can determine the radius of the object $R_{\rm PWN}$ and the energy densities of the ISRF $U_{\rm IR}$ and $U_{\rm OPT}$.
We have already studied the dependence of our calculations on $d$, $U_{\rm IR}$ and $U_{\rm OPT}$ (see Section 2.2 of TT11), and then we do not consider their uncertainties in this paper.
Adopted parameters for each PWN are listed in Table \ref{tbl:parameter}, where we also show results obtained in our previous papers (TeV PWNe) for comparison.

Fitting parameters are the age $\age$, the fraction parameter $\eta$ ($\eta L_{\rm spin}(t)$ is the magnetic power injected from the pulsar), and the parameters of the particle distribution at injection $Q_{\rm inj}(\gamma, t)$ in Equation (\ref{eq_injection}), $\gamma_{\rm max}$, $\gamma_{\rm b}$, $\gamma_{\rm min}$, $p_1$ and $p_2$.
Four of seven parameters, $\gamma_{\rm max}$, $\gamma_{\rm min}$, $p_1$ and $p_2$, determine a broadband spectral shape, and they are usually fitted after the determination of the other three parameters.
Note that only a lower limit of $\gamma_{\rm max}$ and an upper limit of $\gamma_{\rm min}$ can be deduced from observations in radio and X-rays (see Section 2.3 of TT11).
It should also be noted that we do not fit the observed spectral index in X-rays, but fit the flux value.
Because spatial variabilities of the X-ray spectral index are observed for many PWNe, the one-zone spectral model needs not to reproduce the spectral index in X-ray \citep[][TT11]{bet11}.

Three parameters $\age$, $\eta$ and $\gamma_{\rm b}$ are fitted as follows, bearing in mind that typical spectra of young PWNe require $p_1 < 2 < p_2$ and $\eta \ll 1$.
Basically, for young TeV PWNe, three observed values, the power of TeV $\gamma$-rays $P_{\rm IC/ISRF} \propto \gamma_{\rm b} E_{\rm tot}(\age)$, the power of radio and X-rays $P_{\rm syn} \propto \gamma_{\rm b} \eta E^2_{\rm tot}(\age)$ and the spectral break frequency $\nu_{\rm b} \propto \gamma^2_{\rm b} \eta^{1/2} E^{1/2}_{\rm tot}(\age)$ (note that this is not the synchrotron cooling break frequency), can be used to fit $\age$ (or $E_{\rm tot}(\age)$), $\eta$ and $\gamma_{\rm b}$.
However, for non-TeV PWNe, we have only an upper limit in TeV $\gamma$-ray, and then we will get upper limits for $\age$ and $\gamma_{\rm b}$ and a lower limit for $\eta$, i.e., one of these three parameters should be chosen in other ways.
We compare the obtained range of $\age$ with other independent studies about $\age$ (see below).
Note that $E_{\rm tot}(\age)$ is an increasing function of $\age$ and then an uncertainty of $\age$ propagates to $\eta$ and $\gamma_{\rm b}$ as $\eta \propto E^{-4 / 3}_{\rm tot}(\age)$ or $\gamma_{\rm b} \propto E^{1 / 3}_{\rm tot}(\age)$ to retain the observed values of $P_{\rm syn}$ and $\nu_{\rm b}$. 

There are several methods to study the age $\age$ of PWNe.
Observationally, the age $\age$ is estimated from  an expansion rate of PWNe assuming the rate being constant, from  a Doppler shift of emission lines also assuming the expansion velocity being constant, from a proper motion of the central pulsar, and from a possible coincidence of historical events of SNe.
Strictly speaking, although the line emission is from a SN ejecta component, but not from a PWN, this velocity is consistent with the expansion velocity of the PWN for the Crab Nebula.
Theoretically, \cite{c05} studied dynamical evolution (expansion) of PWNe inside a SNR and estimated the age of some PWNe taking account of the spin-down evolution of their central pulsars and of the SN-types, i.e., the density profile of the SN ejecta.
Because these estimates are not always consistent with each other, we just refer to these studies and discuss each object individually.

Once the seven parameters are fitted, then we obtain the expansion velocity $v_{\rm PWN}$, the magnetic field strength $B_{\rm now}$ and the parameters of a central pulsar, the spin-down time $\tau_0$ and the initial rotational energy $L_0 \tau_0$.
Note that we obtain $v_{\rm PWN}$ from $t_{\rm age}$ and the observed extent of a PWN, i.e., our $v_{\rm PWN}$ is independent of the study by C05 described above.
We also get the pair multiplicity inside a pulsar magnetosphere $\kappa$ and the bulk Lorentz factor of a pulsar wind $\Gamma_{\rm w}$ assuming that the pulsar wind is a cold electron-positron plasma flow.
Fitted, and derived parameters for each PWN are also listed in Table \ref{tbl:parameter}.
Note that an uncertainty of $\age$ also propagates to these six derived parameters in the way they decrease with increasing $\age$ except for $L_0 \tau_0$.
Dependence on $\age$ is easy to find from comparing two cases ($\age =$ 2.5kyr and 1kyr) of 3C58 in Table \ref{tbl:parameter}, for example.

\section{3C58}\label{sec:3c58}
3C58 is a well known filled center SNR, and non-thermal radiation is observed in radio, infrared and X-rays.
Although 3C58 has been observed in GeV and TeV $\gamma$-rays, only flux upper limits are obtained.
The angular extent is $\sim 6' \times 9'$ in radio \citep[][]{bet01a, b06}.
A central pulsar of 3C58 (PSR J0205+6449) is observed in radio, X-rays, and $\gamma$-rays with a period $P = 6.57 \times 10^{-2} \rm s$, and a time derivative $\dot{P} = 1.94 \times 10^{-13} \rm s~s^{-1}$ ($\tau_{\rm c} = 5.4 \rm kyr$) \citep[][]{cet02b, met02, let09}. 
We assume $n = 3$.
The age has been estimated with three different methods.
\citet{fet08} showed an average expansion velocity $v_{\rm PWN} \sim 770 \rm km~sec^{-1}$ measured from infrared emission lines.
Corresponding age is $\age \sim 2.5 \rm kyr$ for $R_{\rm PWN} \sim 2 \rm pc$. 
\citet{sg02} argued that 3C58 might be a historical SN in 1181 A.D., i.e., $\age \sim 1 \rm kyr$.
An expansion rate of a radio synchrotron nebula suggests $\age \sim 7 \rm kyr$ \citep[][]{b06} but $\age > \tau_{\rm c}$ is theoretically untenable and cannot be treated in our model because the spin-down time becomes $\tau_0 < 0$.
Moreover, the corresponding expansion velocity is too small as a young PWN.
We thus study the first two cases.
We adopt that the distance to 3C58 is 2 kpc \citep[][]{k10} and then we approximate 3C58 as a sphere of radius $R_{\rm PWN}$ $\sim 2 \rm pc$.
For the ISRF energy densities, we adopt $(U_{\rm IR}, U_{\rm OPT}) = (0.3 \rm eV cm^{-3}, 0.3 \rm eV cm^{-3})$.

\subsection{Results}

%
\begin{figure}
\includegraphics[width=84mm]{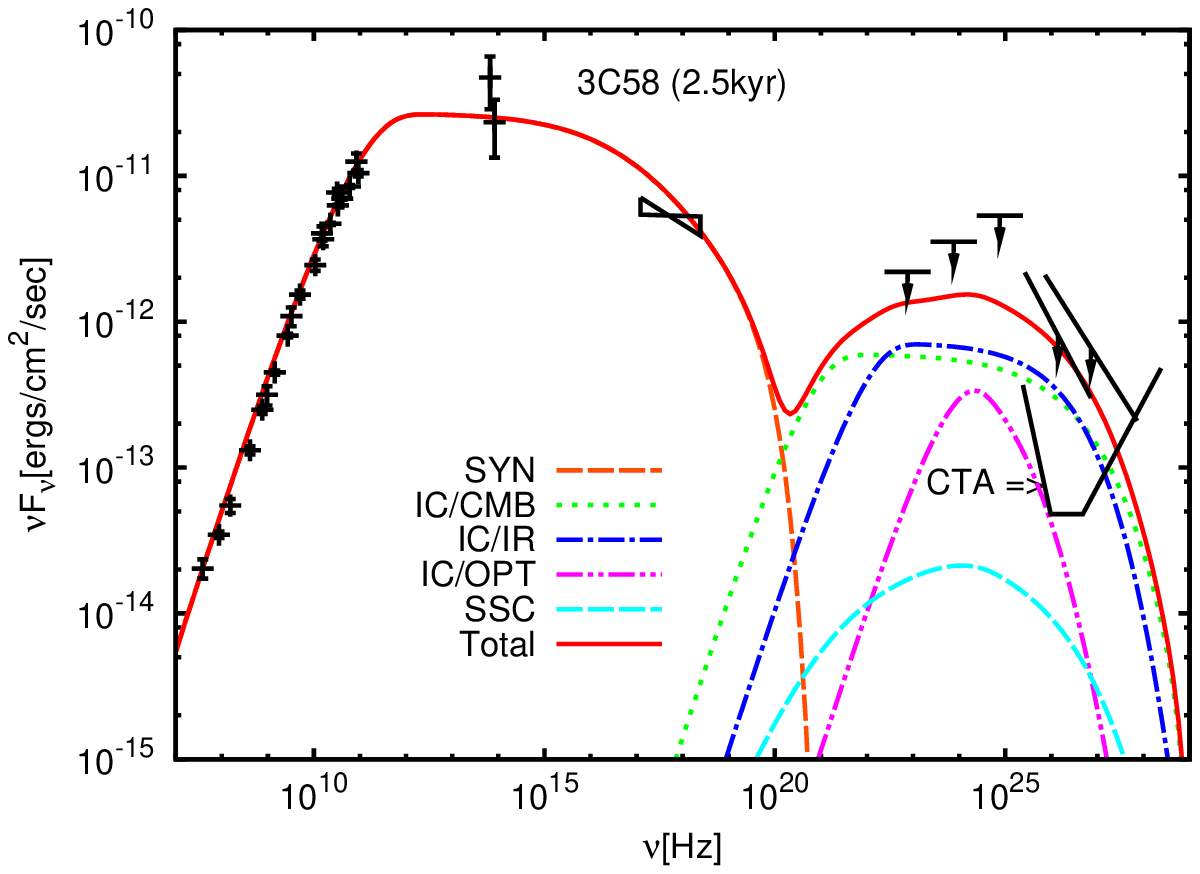}
\caption{
Model spectrum of 3C58 for $\age = 2.5 \rm kyr$.
 The observational data, the upper limits and the sensitivity of CTA (50 hours) are also plotted.
 The observed data are taken from \citet{wet11a, set89, het09, ket06} (radio), \citet{set08} (infrared),  \citet{tet00} (X-ray), \citet{aet11, aet08, aet10} ($\gamma$-ray).
 Used parameters are tabulated in Table \ref{tbl:parameter}.
\label{fig:3c58_1}
}
\end{figure}

First, we show a model spectrum of $\age = 2.5 \rm kyr$ (Figure \ref{fig:3c58_1}).
We reproduce the observed non-thermal spectrum with parameters $\eta = 3.0 \times~10^{-3}$, $\gamma_{\rm max} = 1.0 \times~10^9$, $\gamma_{\rm b} = 9.0 \times~10^4$, $\gamma_{\rm min} = 4.0 \times~10^3$, $p_1 = 1.0$, and $p_2 = 3.0$.
IC/IR and IC/CMB dominate in $\gamma$-rays.
Because the predicted TeV $\gamma$-ray flux in Figure \ref{fig:3c58_1} is almost comparable with the observational upper limit, we can put an upper limit of $\age \lid 2.5 \rm kyr$ and a lower limit of $\eta \gid 3.0 \times~10^{-3}$. 
Accordingly, derived parameters are $L_0 \tau_0 \lid 8.6 \times~10^{48} \rm erg$, $v_{\rm PWN} \gid 780 \rm km~sec^{-1}$, $B_{\rm now} \gid 17 \mu \rm G$ and $\tau_0 \gid 2.9 \rm kyr$.
The fitted, and derived parameters are tabulated in Table \ref{tbl:parameter}.

\begin{figure}
\includegraphics[width=84mm]{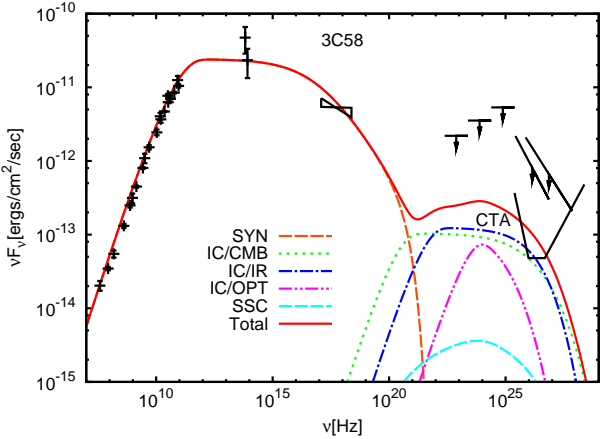}
\caption{
Model spectrum of 3C58 for $\age = 1.0 \rm kyr$.
 The observational data are the same as Figure~\ref{fig:3c58_1}.
 Used parameters are tabulated in Table \ref{tbl:parameter}.
\label{fig:3c58_2}
}
\end{figure} %
Figure \ref{fig:3c58_2} shows a model spectrum with another parameter set being able to reproduce the observations.
For this case, supposing that 3C58 is associated with SN 1181, we take $\age = 1.0 \rm kyr$.
The predicted $\gamma$-ray flux in Figure \ref{fig:3c58_2} is significantly smaller than that in Figure \ref{fig:3c58_1}.
From Table \ref{tbl:parameter}, we can see how the uncertainty of $\age$ propagates to the fitted and derived parameters.
We find the fraction parameter being an order of magnitude larger $\eta =$ $6.0 \times 10^{-2}$ ($B_{\rm now} = 40 \mu \rm G$) and the total injected energy is five times smaller $E_{\rm tot} = 1.1 \times 10^{48} \rm erg$ than those when $\age$ is taken to be 2.5 kyr.
Note that here the fraction parameter does not change as $\eta \propto E^{-4 / 3}_{\rm tot}(\age)$ as was discussed in Section \ref{sec:procedure} because adiabatic and synchrotron coolings of particles are significant for this short age.

\subsection{Discussion}\label{sec:3c58Discussion}
Both spectra in Figures \ref{fig:3c58_1} and \ref{fig:3c58_2} reproduce the observations from radio through X-ray and are compatible with the $\gamma$-ray upper limits.
Although an association with SN 1181 is suggested, some observational \cite[e.g.,][]{fet08} and theoretical (e.g., C05) studies discuss $\age \ga$ 2.0kyr.
In addition to these studies, it is reasonable that the fraction parameter $\eta = 3.0 \times 10^{-3}$ is similar to young TeV PWNe, typically $\eta =$ a few $\times~10^{-3}$ (TT11).
We conclude that 3C58 is about 2.5 kyr old.
For example, the fraction parameter becomes $\eta \sim 10^{-2}$ for $t_{\rm age} = 2.0$ kyr.
The predicted TeV flux for $t_{\rm age} = 2.0$ kyr is about twice as small as that for $t_{\rm age} = 2.5$ kyr (Figure \ref{fig:3c58_1}).
Future observations in TeV $\gamma$-rays will distinguish the models.

Although past dynamical and spectral evolution studies of 3C58 also obtained a similar age of $2.4 \rm kyr$ (C05) and $2 \rm kyr$ \citep[][]{bet11}, they adopt different distance $d = 3.2 \rm kpc$ while we adopt $d = 2 \rm kpc$ taken from \citet{k10}.
Despite this difference, we try to compare our model with the study by \citet{bet11}.
They obtained $\age \sim 2 \rm kyr$ and $B(\age) \sim 43 \mu \rm G$.
Because the magnetic field strength is determined from the observed ratio of  synchrotron and inverse Compton scattering $P_{\rm syn} / P_{\rm IC/ISRF}$, it does not depend on the distance to 3C58.
Their magnetic field strength $\sim 43 \mu \rm G$ is consistent within our condition $B_{\rm now}$ $\gid 17 \mu \rm G$ and $P_{\rm IC/ISRF}$ is similar to Figure \ref{fig:3c58_2} ($B_{\rm now} =$ $40 \mu \rm G$).

Radio observations of 3C58 suggested a flux evolution.
\citet{iet04} found non-stationary variations of the flux densities at 1.5 $-$ 8.5 GHz.
Although they found a frequency independent increase of the radio flux $\sim 0.3 \% \rm yr^{-1}$ during the period 1965 $-$ 1986, they also found a frequency dependent variation during the period 1966 $-$ 2003, where both increase and decrease of fluxes are found ranging from -0.3 to 0.5 $\% \rm yr^{-1}$.
Such a chaotic behavior is different from the case of the Crab Nebula \citep[][]{v07} and is unexpected as a PWN.
On the other hand, a recent observed rate of variabilities by \citet{wet11a} (seven-year {\it WMAP} observations) seems to be zero within the margin of error, $- 0.05 \pm 0.09 \% \rm yr^{-1}$ at 23GHz, $0.14 \pm 0.15 \% \rm yr^{-1}$ at 33GHz, $- 0.17 \pm 0.18 \% \rm yr^{-1}$ at 41GHz, $- 0.43 \pm 0.35 \% \rm yr^{-1}$ at 61GHz, and $0.47 \pm 0.80 \% \rm yr^{-1}$ at 94GHz.
In our model ($\age = 2.5 \rm kyr$), the current decrease rates are about $- 0.04 \% \rm yr^{-1}$ at 1GHz, $- 0.05 \% \rm yr^{-1}$ at 10GHz, $- 0.06 \% \rm yr^{-1}$ at 100GHz.
They are almost independent of frequencies and are a factor of four smaller than the case of the Crab Nebula.
Because the absolute radio flux of 3C58 is also two orders of magnitude less than that of the Crab Nebula, we conclude that such a small rate of decrease is difficult to detect and that our result is consistent with {\it WMAP} observations.

The high energy power-law index at injection $p_2 = 3.0$ for 3C58 is different from young TeV PWNe, for which typically $p_2 \sim 2.5$ (TT11).
When we use $p_2 \sim 2.5$, the calculated X-ray flux becomes almost an order of magnitude larger than the observed X-ray flux by \citet{tet00}.
This result does not change even if we ignore the {\it Spitzer} observation in infrared.

\section{Other Young Non-TeV PWNe}\label{sec:other}

We also study other four young non-TeV PWNe, G310.6-1.6, G292.0+1.8, G11.2-0.3 and B0540-69.3.
We found that their flux upper limits in TeV $\gamma$-rays only weakly constrain the fitting parameters.
For example, we obtain a lower limit of $\eta \gid 5 \times 10^{-4}$ for G310.6-1.6 and lower limits of $\eta$ are much lower than $10^{-5}$ for other three objects.
Hence, $\eta$ being a few $\times~10^{-3}$ is also consistent for these four PWNe as well as 3C58 and young TeV PWNe.
We regard that $\eta \sim$ a few $\times~10^{-3}$ is reasonable for all of young PWNe. 
Figures \ref{fig:g310} $-$ \ref{fig:b0540} show model spectra of G310.6-1.6, G292.0+1.8, G11.2-0.3 and B0540-69.3 when the fraction parameters are taken to be $\eta = 3.0 \times 10^{-3}$ together with observational data.
Adopted, fitted, and derived values corresponding to Figures \ref{fig:g310} $-$ \ref{fig:b0540} are listed in Table \ref{tbl:parameter}.
Below, we briefly discuss each object.

\subsection{G310.6-1.6}\label{sec:g310}

%
\begin{figure}
\includegraphics[width=84mm]{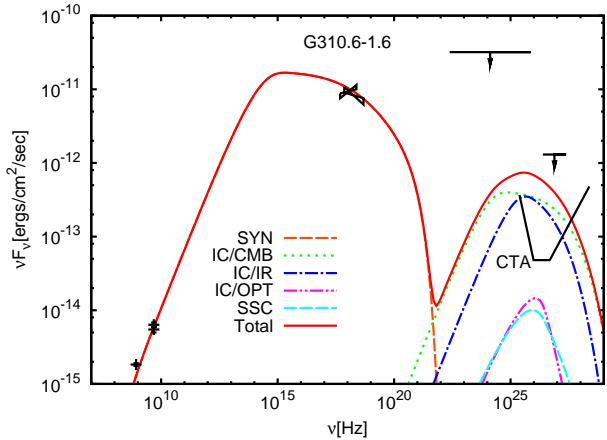}
\caption{
Model spectrum of G310.6-1.6.
The observational data, the upper limits and the sensitivity of CTA (50 hours) are also plotted.
The observed data are taken from \citet{ret10} (radio, X-ray and $\gamma$-ray).
Used parameters are tabulated in Table \ref{tbl:parameter}.
\label{fig:g310}
}
\end{figure}

G310.6-1.6 is a recently discovered composite SNR at $d = 7 \rm kpc$, and the associated pulsar (PSR J1400-6325) has $L_{\rm spin} = 5.1 \times 10^{37} \rm erg~s^{-1}$ and $\tau_{\rm c} = 12.7 \rm kyr$ ($n = 3$) \citep[][]{ret10}.
A model spectrum with observations is shown in Figure \ref{fig:g310}.
\citet{ret10} discussed a one-zone leptonic model of G310.6-1.6, and they obtained conditions of the magnetic field ($> 6 \mu \rm G$) and the age ($< 1.9 \rm kyr$) from a TeV $\gamma$-ray flux upper limit.
Our results are compatible with their conditions.

For $\eta = 3.0 \times 10^{-3}$ ($B_{\rm now} = 6.7 \mu \rm G$), the age becomes $\age =$ 0.6 kyr and then G310.6-1.6 is one of the youngest PWNe.
We consider that a larger $\eta$,i.e., a larger $v_{\rm PWN}$, is unlikely because $v_{\rm PWN} \sim 2120 \rm km~sec^{-1}$ for $\age =$ 0.6 kyr is already large compared with other young PWNe. 
Moreover, \citet{ret10} suggested $\age \la 1 \rm kyr$ based on some theoretical studies of dynamical evolutions of a composite SNR.
Our result of $\age =$ 0.6 kyr shows that the predicted TeV $\gamma$-rays from G310.6-1.6 is about three times smaller than the {\it HESS} upper limit.

\subsection{G292.0+1.8}\label{sec:g292}

%
\begin{figure}
\includegraphics[width=84mm]{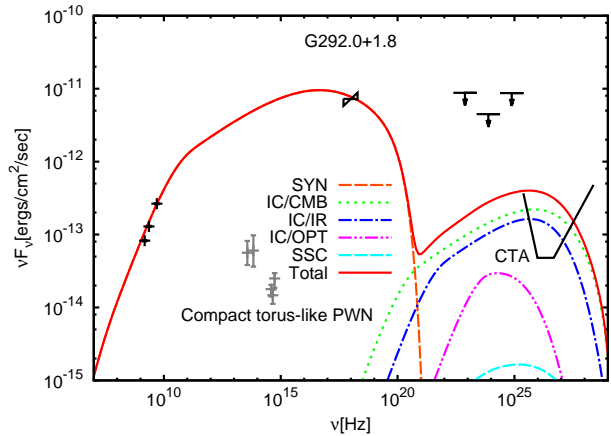}
\caption{
Model spectrum of G292.0+1.8.
The observational data, the upper limits and the sensitivity of CTA (50 hours) are also plotted.
The observed data are taken from \citet{gw03} (radio), \citet{zet09} (infrared), \citet{zet08} (optical), \citet{het01} (X-ray) and \citet{aet11} ($\gamma$-ray).
Used parameters are tabulated in Table \ref{tbl:parameter}.
\label{fig:g292}
}
\end{figure}

G292.0+1.8 is a well-known Cas A-like (Oxygen-rich) SNR \citep[][]{pet07, get12} and has the central pulsar (PSR J1124-5916) with $L_{\rm spin} = 1.1 \times 10^{37} \rm erg~s^{-1}$ and $\tau_{\rm c} = 2.9 \rm kyr$ ($n = 3$) \citep[][]{cet02a}.
We adopt $d = 6 \rm kpc$ \citep[][]{gw03}.
A model spectrum with observations is Figure \ref{fig:g292}.
Although optical and infrared observations found a compact torus-like feature around PSR J1124-5916 \citep[][]{zet08, zet09}, these do not seem to account for the total PWN emission at these wavelengths (plotted as a lower limit in Figure \ref{fig:g292}).
This is because the detected structure is much smaller than radio and X-ray nebula.

No other broadband spectral study of G292.0+1.8 is found in the literature to our knowledge.
Although C05 obtained $\age \sim 3.2 \rm kyr$ and \citet{wet09} estimated $\age \sim 3.0 \rm kyr$ from a proper motion of oxygen-filaments, $\age > \tau_{\rm c}$ is forbidden with $n = 3$ (C05 considered the case $n < 3$).
Our result is $\age = 2.7 \rm kyr \la \tau_{\rm c}$.
Although the values of $\age$ and $n$ are somewhat different between our and C05's models, the internal energy inside the PWN $E_{\rm int}(\age) \sim 2.1 \times 10^{48} \rm erg$ for our model is similar to theirs $E_{\rm int}(\age) = (1 - 2) \times 10^{48} \rm erg$, i.e.,  PSR J1124-5916 has released similar amount of its rotational energy for both models.
We consider that our result is consistent with C05 and \citet{wet09}.
Lastly, it should be noted that we do not need the low-energy component of the injection spectrum $Q_{\rm inj}$ similar to the case of G0.9+0.1 studied in TT11 (radio-emitting particles are created by adiabatic cooling of particles of $\gamma \gid \gamma_{\rm b}$).

\subsection{G11.2-0.3}\label{sec:g11.2}

%
\begin{figure}
\includegraphics[width=84mm]{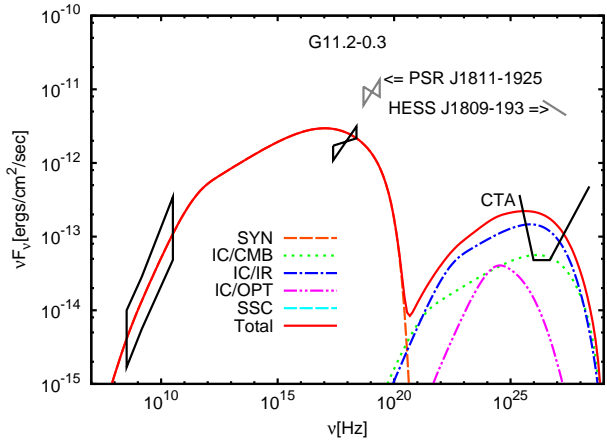}
\caption{
Model spectrum of G11.2-0.3.
The observational data, the upper limits and the sensitivity of CTA (50 hours) are also plotted.
The observed data are taken from \citet{tet02} (radio), \citet{ret03} (X-ray), \citet{det08} (X-ray from PSR J1811-1925) and \citet{aet07} ($\gamma$-ray from HESS J1809-193).
Used parameters are tabulated in Table \ref{tbl:parameter}.
\label{fig:g11.2}
}
\end{figure}

G11.2-0.3 is a composite SNR at $d = 5 \rm kpc$ \citep[][]{bet85, get88}, and the associated pulsar (PSR J1811-1925) has $L_{\rm spin} = 6.8 \times 10^{36} \rm erg~s^{-1}$ and $\tau_{\rm c} = 23 \rm kyr$ ($n = 3$) \citep[][]{tet99, get04}.
A model spectrum with observations is shown in Figure \ref{fig:g11.2}.
\citet{aet07} found a TeV $\gamma$-ray source (HESS J1809-193) in the vicinity of G11.2-0.3.
Although some discussions about an association between them have been made, their association is considered unlikely by \citet{kp07, det08}.
We conclude that they are not associated from Figure \ref{fig:g11.2}.
We also confirm that the hard X-ray flux at $\sim 10^{19} \rm Hz$ is not explained by the PWN emission as \citet{det08}.
They concluded that it is a contribution from PSR J1811-1925.

No other broadband spectral study of G11.2-0.3 is found in the literature to our knowledge.
Despite the large $\tau_{\rm c}$ of PSR J1811-1925, G11.2-0.3 is thought as a young SNR/PWN system because of its almost spherical morphology of the PWN and a small off-set of the pulsar from the center of the SNR.
Moreover, G11.2-0.3 is suspected to be associated with a historical SN in A.D. 386, i.e., $\age \sim 1.6 \rm kyr$ \citep[][]{cs77} and observations of expansion rates in radio \citep[][]{tr03} and in infrared \citep[][]{ket07} support their association.
C05 obtained $\age \sim 1.6 \rm kyr$.
Our result $\age = 2.0 \rm kyr$ is consistent with these age estimates. 

\subsection{SNR B0540-69.3}\label{sec:b0540}

%
\begin{figure}
\includegraphics[width=84mm]{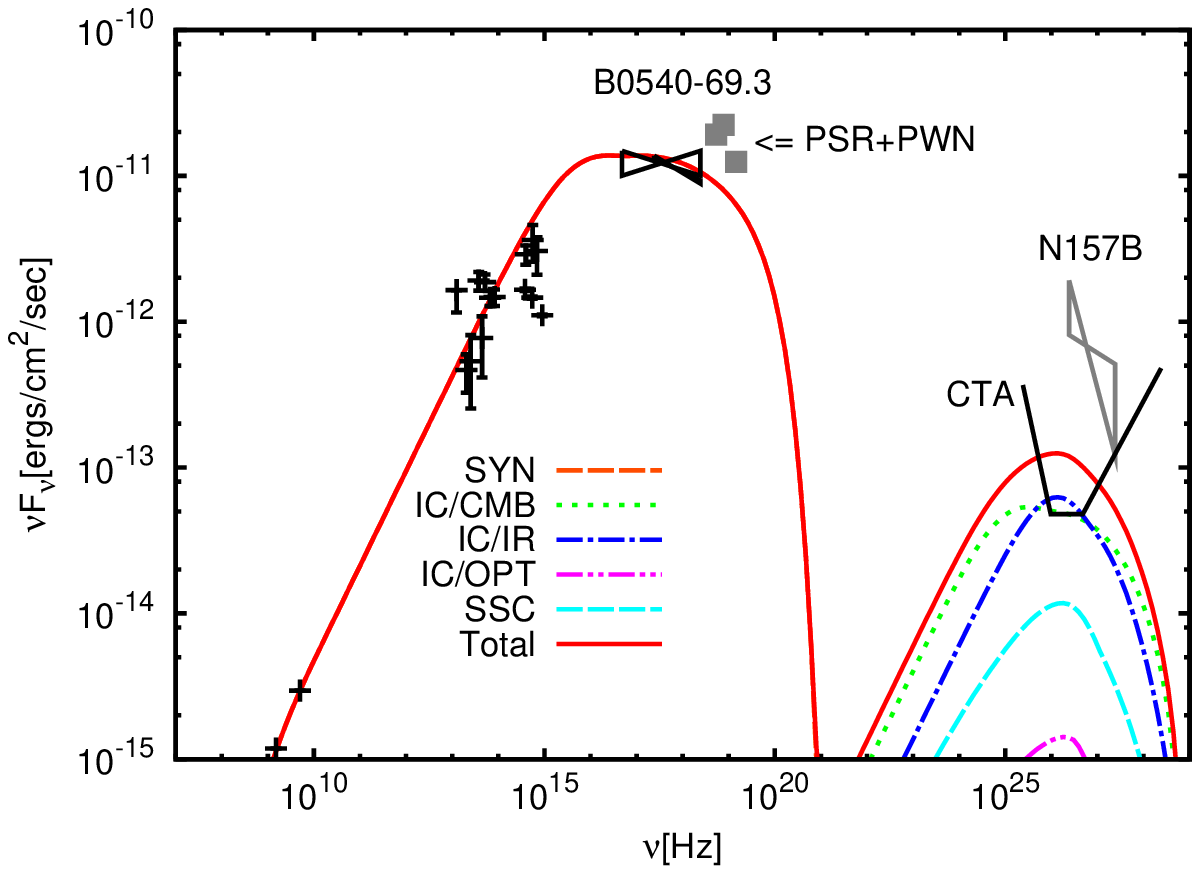}
\caption{
Model spectrum B0540-69.3.
The observational data, the upper limits and the sensitivity of CTA (50 hours) are also plotted.
The observed data are taken from \citet{met93} (radio), \citet{gt98, set04} (infrared and optical), \citet{het02, ket01} (X-ray), \citet{cet08} (combined flux of PSR and PWN), and \citet{ket12} ($\gamma$-ray from N157B).
Used parameters are tabulated in Table \ref{tbl:parameter}.
\label{fig:b0540}
}
\end{figure}

B0540-69.3 is a composite SNR in the Large Magellanic Cloud (LMC, $d \sim 50 \rm kpc$), and the associated pulsar (PSR B0540-69) has $L_{\rm spin} = 1.48 \times 10^{38} \rm erg~s^{-1}$, $\tau_{\rm c} = 1.7 \rm kyr$ and $n \sim 2.1$ \citep[][]{get11, let05}.
Because the ISRF in LMC is uncertain, we tentatively use values around the Sun $(U_{\rm IR}, U_{\rm OPT}) = (0.5 \rm eV, 0.5 \rm eV)$.
A model spectrum with observations is shown in Figure \ref{fig:b0540}.
Although \citet{ket12} detected TeV $\gamma$-rays from N157B close to B0540-69.3, B0540-69.3 was not detected.
\citet{cet08} showed a combined flux of the PWN and pulsar in hard X-rays and argued that the PWN emission contributes about 75 $-$ 80 \%.
Our result seems consistent with \citet{cet08}.
Although we fail to fit infrared and optical observations, these fluxes seem to have considerable uncertainties because a spatially varying background makes background subtraction uncertain as discussed by \citet{wet08}.

No other broadband spectral study of B0540-69.3 is found in the literature to our knowledge.
C05 obtained $\age \sim 0.8 \rm kyr$ which is consistent with our calculation.
\citet{met06} found a line-width of filaments $\sim 3300 \rm km~sec^{-1}$ (corresponding expansion speed is a half of this value) which is also consistent with our calculation.
However, they conclude $\age \sim 1.2 \rm kyr$ because faint materials extend ($\sim$ 8'') slightly larger than the PWN $\sim$ 5''.
We consider that $\age \sim 0.8 \rm kyr$ in our model is acceptable.
If the ISRF around B0540-69.3 is close to or larger than the assumed values $(U_{\rm IR}, U_{\rm OPT}) = (0.5 \rm eV, 0.5 \rm eV)$, B0540-69.3 may be detected by future $\gamma$-ray observations.

\section{DISCUSSIONS}\label{sec:discussion}

In Sections \ref{sec:3c58} and \ref{sec:other}, we found that all the young PWNe can be fitted by a similar value of the fraction parameter $\eta \sim$ a few $\times~10^{-3}$.
\citet{bet11} also studied spectral evolution of young PWNe, but their parameter corresponding to $\eta$ in our model takes a broad range of values.
This is because a magnetic evolution in their model is different from ours (Equation (\ref{eq:magnetic-field})).
Both their and our models are not rigorous treatments based on magnetohydrodynamic equations, and then we do not discuss which is a better approximation.
However, because we presume it plausible that properties of all young PWNe are similar to each other, in the discussion below, we do not consider the case when $\eta$ significantly differs from a few $\times~10^{-3}$, i.e., the following discussions are based on the parameters in Table \ref{tbl:parameter}.
We especially focus on the differences and similarities of TeV and non-TeV PWNe.

\begin{figure}
\includegraphics[width=84mm]{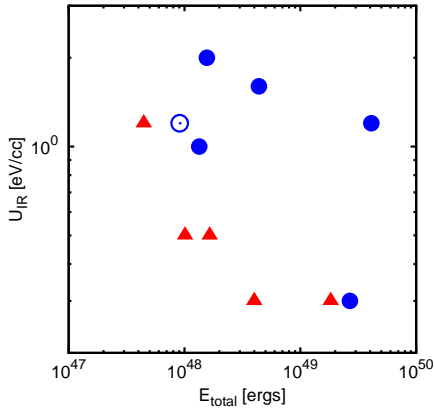}
\caption{
The correlation between the total injected energy $E_{\rm tot}$ versus the ISRF energy density in infrared $U_{\rm IR}$ of young PWNe.
TeV PWNe are shown by filled circles except for Kes 75 (open circle) and non-TeV PWNe are by filled triangles.
Young TeV PWNe are located on the top-right region, and young non-TeV PWNe are located on the bottom-left region.
\label{fig:detectability}
}
\end{figure}

We first discuss the detectability of TeV $\gamma$-rays from young PWNe.
We search for correlations between all the parameters and the detectability in TeV $\gamma$-rays.
Figure \ref{fig:detectability} shows the correlation between the total injected energy $E_{\rm tot}$ versus the ISRF energy density in infrared $U_{\rm IR}$.
TeV and non-TeV PWNe are shown by circles and triangles, respectively.
As seen from Figure \ref{fig:detectability}, TeV PWNe are located on the top right region (large values of $E_{\rm tot}$ and $U_{\rm IR}$) and non-TeV PWNe are located on the bottom left region (small values of $E_{\rm tot}$ and $U_{\rm IR}$).
This behavior indicates the importance of total particle energy stored in PWNe and the energy density of target photons of inverse Compton scattering for the detectability of TeV $\gamma$-rays from young PWNe.
As claimed by \citet{h07}, the local ISRF energy density is important for the detectability of TeV $\gamma$-ray.
The Crab Nebula (the lowermost circle in Figure \ref{fig:detectability}) is in the lowest ISRF environment, although its $\gamma$-ray emission is dominated by SSC component.
G11.2-0.3 (the uppermost triangle in Figure \ref{fig:detectability}) is TeV-undetected although it is located in a high ISRF environment.
The positions of these two objects in Figure 7 suggest that the local ISRF alone does not determine the TeV $\gamma$-ray detectability.
For other combinations of the parameters, such as the distance $d$ (determining the observed flux), the radius $R_{\rm PWN}$ (determining the degree of adiabatic cooling), or $\tau_{\rm c}$ and $L_{\rm spin}$ \citep[studied by][]{met09}, we do not find any significant correlations, and we do not show them here.

\begin{figure}
\includegraphics[width=84mm]{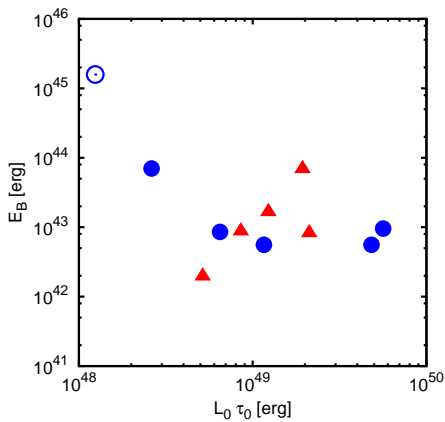}
\caption{
The correlation between the initial rotational energy $L_0 \tau_0$ versus the magnetic energy $B^2_{\ast}R^3_{\ast} / 6$ of the central pulsars.
TeV PWNe are filled circle except for Kes 75 (open circle) and non-TeV PWNe are filled triangle.
}
\label{fig:rot-mag}
\end{figure}

The properties of central pulsars may be different from TeV to non-TeV young PWNe.
Following TT11, we characterize each pulsar with the initial rotational energy and the magnetic energy.
In Figure \ref{fig:rot-mag}, we plot the correlation between the initial rotational energy $L_0 \tau_0$ versus the magnetic energy of the pulsar $E_{\rm B} = B^2_{\ast}R^3_{\ast} / 6$, where $B_{\ast} \propto \dot{P}^{1/2} P^{1/2}$ (assuming magnetic dipole radiation) and $R_{\ast} = 10^6 \rm cm$ are the surface dipole magnetic field and the radius of the pulsar, respectively.
We find no significant difference between the central pulsars of young TeV and non-TeV PWNe.
Except for Kes 75, the initial rotation and the magnetic energies are uniformly distributed between $L_0 \tau_0 \sim 10^{48 - 50} \rm erg$ and $E_{\rm B}\sim 10^{42 - 44} \rm erg$ ($2 \times 10^{12} - 3 \times 10^{13} \rm G$ for the surface dipole magnetic field strength).

For the parameters of a particle distribution at injection $Q_{\rm inj}(\gamma, t)$, while we found that $p_2 = 2.5$ is common to TeV PWNe, we find no typical value for each parameter anymore when we include non-TeV PWNe.
It may be because that spatially varying X-ray spectral indices which are observed from most PWNe do not guarantee a fitting by a one-zone model.
It is not clear why TeV PWNe can be fitted by a common value of $p_2 = 2.5$. 
On the other hand, radio and optical emissions are more spread in entire PWNe than X-ray, i.e., showing approximately one-zone behaviors, and then the varieties of $p_1$ and $\gamma_{\rm b}$ seems real.
Interestingly, G292.0+1.8 does not need the low-energy power-law component at injection as TeV PWN G0.9+0.1 also shows the same characteristic (TT11), but we do need the low-energy component for other young PWNe.
Lastly, following TT11, we can estimate a lower limit of the multiplicity $\kappa$ and a upper limit of the bulk Lorentz factor $\Gamma_{\rm w}$ of a pulsar wind (listed in Table \ref{tbl:parameter}).
These parameters are also common to TeV and non-TeV PWNe.

\section{CONCLUSIONS}\label{sec:conclusion}

We applied the spectral evolution model developed by TT11 to young non-TeV PWNe, 3C58, G310.6-1.6, G292.0+1.8, G11.2-0.3 and B0540-69.3 to study differences between non-TeV and TeV PWNe.
Observed TeV $\gamma$-ray flux upper limits of these non-TeV PWNe give us only upper or lower limits of the model parameters.
However, we can estimate plausible values of parameters from other observational or theoretical studies which are independent off our calculations.
For the case of 3C58, the fitting parameters are significantly constrained and a lower limit of the fraction parameter is $\eta \gid 3.0 \times 10^{-3}$ with $t_{\rm age} \lid 2.5 \rm kyr$.
Combined with the age estimated from the expansion velocity of filamentary structures, we consider that $\eta \sim 3.0 \times 10^{-3}$ with $\age \sim 2.5 \rm kyr$ is reasonable for 3C58.
This value of fraction parameter is similar to that of young TeV PWNe ($\eta \sim$ a few $\times~10^{-3}$).
For other young non-TeV PWNe, we also find that the fraction parameters of $\eta \sim$ a few $\times~10^{-3}$ is consistent with observations although a wide range of values is nominally allowed.
Our model alone weakly constrains the fraction parameters with a smaller value of a lower limit.
The comparison between other observational and theoretical studies makes the fraction parameters of young PWNe being $\eta \sim$ a few $\times~10^{-3}$ acceptable.

For the detectability of young PWNe in TeV $\gamma$-rays, we find the detectability is correlated with both the total injected energy $E_{\rm tot}$ and the ISRF energy density in infrared $U_{\rm IR}$.
This behavior indicates importance of the total particle energy stored in PWNe and the energy density of target photons of inverse Compton scattering for the detectability of TeV $\gamma$-rays from young PWNe.
We do not find any significant correlations for other combinations of the parameters.

The central pulsar properties, i.e., the initial rotational energy $L_0 \tau_0$ and the magnetic energy $E_{\rm B}$ of the pulsar, are not significantly different between TeV and non-TeV PWNe.
The values are distributed in a range of $L_0 \tau_0 \sim 10^{48 - 50} \rm erg$ and $E_{\rm B} \sim 10^{42 - 44} \rm erg$. 

The fitted break energy $\gamma_{\rm b}$ and the high energy power-law index at injection $p_2$ are widely distributed, $\gamma_{\rm b} \sim 10^{4 - 7}$ and $p_2 \sim 2.5 - 3.0$.
Interestingly, G292.0+1.8 does not need the low-energy component at injection to reproduce the observed radio spectrum.
This behavior is the same as G0.9+0.1 studied in the previous paper, i.e., the adiabatic cooling of the high-energy component reproduces radio emissions from the object.

The radio flux evolution of 3C58 calculated in our model is about $- 0.04 \% \rm yr^{-1}$ which is four times smaller than that of the Crab Nebula.
3C58 is the only one PWN whose radio flux evolution are obtained, except for the Crab Nebula.
The latest {\it WMAP} observation shows that the flux evolution is almost zero within error and then we conclude that the radio flux evolution calculated in our model is consistent with the {\it WMAP} observation.

\section*{Acknowledgments}

This work is partly supported by JSPS Research Fellowships for Young Scientists (S. T., 23250) and KAKENHI (F. T., 20540231).

\appendix

\section{SPECTRAL EVOLUTION MODEL OF YOUNG PWNE}\label{app:model}
We have modeled spectral evolution of young PWNe.
The radiation processes are synchrotron radiation and inverse Compton scattering off synchrotron radiation (SSC) and off the ISRF (IC/ISRF).
The ISRF has three components (the cosmic microwave background radiation (CMB), infrared photons from dust grains and optical photons from stars).
The infrared and optical components of the ISRF are parametrized by the energy densities $U_{\rm IR}$ and $U_{\rm OPT}$ which vary with the distance from the Galactic center $r$ and the height above the Galactic plane $z$.

We consider a PWN which is a uniform sphere expanding at a constant velocity $v_{\rm{PWN}}$ (the radius of the PWN $R_{\rm PWN}(t) = v_{\rm PWN} t$).
The PWN contains a magnetic field and a population of accelerated electron-positron plasma, which are injected from its central pulsar.
The spin-down power $L_{\rm spin}(t)$ of the central pulsar is divided into the magnetic power $\eta L_{\rm spin}(t)$, and the particle power $(1 - \eta) L_{\rm spin}(t)$, where $\eta$ ($0 \lid \eta \lid 1$) is the fraction parameter.
Evolution of the spin-down power is given by $L_{\rm spin}(t) = L_{\rm 0} [1+(t/\tau_0)]^{-(n+1)/(n-1)}$, where $\tau_0$ is the spin-down time and $L_0 \tau_0$ is the initial rotational energy.
We need four values to specify $L_{\rm spin}(t)$, the current pulsar period $P$, its time derivative $\dot{P}$, braking index $n$ and the current age $\age$.
The characteristic age $\tau_{\rm c} \equiv P / 2 \dot{P}$ is described as $\tau_{\rm c} = (n - 1)/2 \cdot (\tau_0 + \age)$.

For evolution of a particle distribution $N(\gamma, t)$, we use the continuity equation in the energy space,
\begin{equation}\label{eq_continuity}
\frac{\partial}{ \partial t} N(\gamma, t) + \frac{ \partial}{\partial \gamma} \left( \dot{\gamma}(\gamma, t) N(\gamma, t) \right) = Q_{\mathrm{inj}}(\gamma, t).
\end{equation}
The particle injection $Q_{\rm inj}(\gamma, t)$ is assumed to follow a broken power-law characterized by five parameters $\gamma_{\rm min}$, $\gamma_{\rm b}$, $\gamma_{\rm max}$, $p_{\rm 1}$ and $p_{\rm 2}$, which are the minimum, break and maximum Lorentz factors and the power-law indices at the low and high energy ranges of the injection spectra, respectively.
The particle injection $Q_{\rm inj}(\gamma, t)$ has a form of
\begin{equation}\label{eq_injection}
Q_{\rm inj}(\gamma, t) = \left\{
\begin{array}{ll}
Q_{\rm 0}(t) (\gamma / \gamma_{\rm {b}})^{-p_{\rm 1}} & \mbox{ for $\gamma_{\rm min} \lid \gamma \lid \gamma_{\rm b}  $ ,} \\
Q_{\rm 0}(t) (\gamma / \gamma_{\rm {b}})^{-p_{\rm 2}} & \mbox{ for $\gamma_{\rm b}   \lid \gamma \lid \gamma_{\rm max}$ ,}
\end{array} \right.
\end{equation}
where $\gamma$ is the Lorentz factor of relativistic particles.
We require that the normalization $Q_{\rm{0}}(t)$ satisfies $(1-\eta)L(t) = \int_{\gamma_{\rm min}}^{\gamma_{\rm max}} Q_{\rm inj}(\gamma, t) \gamma m_{\rm e} c^{2} d\gamma$, where $m_{\rm e}$ and $c$ are the mass of an electron (or positron) and the speed of light, respectively.
The cooling term $\dot{\gamma}(\gamma, t)$ includes the radiative coolings (synchrotron radiation and IC/ISRF) and the adiabatic cooling.
Lastly, we need to specify the magnetic field evolution, which is assumed to be given by magnetic energy conservation,
\begin{eqnarray}\label{eq:magnetic-field}
\frac{4\pi}{3} (R_{\rm{PWN}}(t))^{3} \cdot \frac{(B(t))^{2}}{8\pi} & = & \int_0^{t} \eta L(t') dt' \nonumber \\
& = & \eta E_{\rm tot}(t),
\end{eqnarray}
where $E_{\rm tot}(t)$ is the integrated spin-down power at a time $t$.
Although Equation (\ref{eq:magnetic-field}) neglects adiabatic losses, the magnetic field also suffers from winding and stretching process.
Uncertainties of one-zone models may be larger than these corrections.
Note that some justifications of Equation (\ref{eq:magnetic-field}) are discussed in Section 2.2 of \citet{tt10}.

\bsp

\label{lastpage}

\clearpage

\end{document}